\documentclass[preliminary,copyright,creativecommons]{eptcs}

\providecommand{\event}{GASCom 2024} 

\usepackage{iftex}
\usepackage[usestackEOL]{stackengine}
\usepackage{graphicx}
\usepackage{multirow}
\usepackage{calligra}
\usepackage{latexsym}
\usepackage{amssymb}
\usepackage{amsfonts}
\usepackage{amsmath}
\usepackage{indentfirst}

\ifpdf
  \usepackage{underscore}         
  \usepackage[T1]{fontenc}        
\else
  \usepackage{breakurl}           
\fi

\newtheorem{definition}{Definition}[section]

\title{Dyck Paths Enumerated by the $\mathbb Q$-bonacci Numbers}

\author{Elena Barcucci\qquad
Antonio Bernini\qquad
Stefano Bilotta\qquad
Renzo Pinzani
\institute{Dipartimento di Matematica e Informatica ``U. Dini''}
\institute{Universit\`a degli
Studi di Firenze\\
Viale G.B. Morgagni 65, 50134 Firenze, Italy.}
\email{\Longunderstack[l]
{elena.barcucci@unifi.it
\\antonio.bernini@unifi.it
\\stefano.bilotta@unifi.it
\\renzo.pinzani@unifi.it}}
}

\def\titlerunning{Dyck Paths Enumerated by the $\mathbb Q$-bonacci Numbers}
\def\authorrunning{E. Barcucci, A. Bernini, S. Bilotta \& R. Pinzani}
\begin{document}

\maketitle
\section{Introduction}
The $q$-generalized Fibonacci numbers \cite{Fei,Mil} can be combinatorially interpreted in many different ways, for $q\in\mathbb N^+$.
One of them involves the Dyck paths having height at most two. More precisely, the subsets of these bounded Dyck paths avoiding $q$ consecutive valleys at height $1$ are enumerated according to their semilength (i.e. the     number of steps of the path divided by 2) by the mentioned sequence \cite{BBP1, BBP2}.   
The same numbers count the binary strings avoiding $q+1$ consecutive $1$'s, with $q\geq 1$, according to their length \cite{Ber, Vaj}. Note that to be exact we should write that the enumerating sequences are the $(q+1)$-generalized Fibonacci numbers, for $q\geq 1$. 

Not long ago Baril, Kirgizov, and Vajnovszki \cite{BKV} introduced the set $\mathcal W_n^q$ of the $q$-decreasing strings which are binary strings of length $n$ where each maximal factor $0^a1^b$, with $a>0$, satisfies $q\cdot a>b$, for $q\geq 1$. Moreover, among other results, the authors give a bijection between $\mathcal W_n^q$ and the set $\mathcal B_n(1^{q+1})$ of the binary strings of length $n$ avoiding $q+1$ consecutive $1$'s, for $q\geq 1$.  

Recently \cite{Kir}, Kirgizov generalized the $q$-decreasing strings to the case where $q$ is a positive rational number, $q\in\mathbb Q^+$, and they are enumerated by the numbers called $\mathbb Q$-bonacci by the author. 
In this paper we provide a class of restricted Dyck paths that result in having the same enumeration (according to their semilength). More precisely, we consider the Dyck paths having height at most two and introduce some constraints on the numbers of consecutive valleys at height one which must be followed by a suitable number of valleys at height zero, depending on the value of $q\in\mathbb Q^+$. 

\section{Preliminaries}

In the paper we indicate a Dyck path in linear notation as a string over the alphabet $\{U,D\}$, where $U$ and $D$ replace the up and the down steps of the path, respectively. We figure a Dyck path in a Cartesian coordinate system, starting from the origin and ending in a point of the $x$-axis. A valley is a substring $DU$, while a peak is a substring $UD$. The height of a valley is the ordinate reached by the $D$ step. We refer to a valley at height 1 or at height $0$ with \emph{1--valley} or  \emph{0--valley}, respectively.
For the peaks, with \emph{1--peak} (or \emph{0--peak}) we mean a peak $UD$ whose $D$ step reaches the ordinate $1$ (or $0$).  

Since we are going to deal only with Dyck paths having height at most 2, we do not record this restriction in our notation.

Let $\mathcal D_n^q$, with $q\geq 1$, denote the set of the Dyck paths having height at most 2, and avoiding $q+1$ consecutive 1--peaks (which is the same as avoiding $q$ consecutive 1--valleys), and having semilength $n$, where $n\geq 0$.

Clearly, a Dyck path $P\in \mathcal D_n^q$
starts with one of the factors $UD, UUDD,UUDUDD,$
$\ldots, U(UD)^qD$ (where $(UD)^q$ is the string obtained by concatenating $UD$ to itself $q$ times).
Then the path $P$ is obtained by concatenating one of these factors to a path of suitable length, and the set $\mathcal D_n^q$ can be generated by
$$
\mathcal D_n^q = \begin{cases}
\varepsilon, & \text{if $n=0$}; \\
\displaystyle\bigcup_{j=0}^qU(UD)^{j}D\cdot \mathcal D_{n-1-j}^{q}, & \text{if $n\geq 1$.}
\end{cases}
$$
It is thus enumerated by the sequence of the $(q+1)$-generalized Fibonacci numbers
\begin{equation*}
f_n^{(k)}=
\begin{cases}
1,  & \text{if $n=0$};\\
\displaystyle\sum_{i=1}^{k}f_{n-i}^{(k)}, & \text{if $n\geq 1
\quad (f_{\ell}^{(k)}=0$}  \text{ if $\ell<0$}).
\end{cases}\quad 
\end{equation*}
Note that for $q=1$ we get the classical Fibonacci numbers.

\section{Construction in the case $q\in \mathbb Q^+$}
\subsection{The particular case $q=1/s$}
If $P\in \mathcal D_n^q$ with $q\in \mathbb N^+$, then $q$ consecutive $1$--peaks in $P$ are necessarily followed by at least one $0$--valley.
In the case where $q\in \mathbb Q^+$, we require that the number of $0$--valleys have some constraints. 

\smallskip
We start with the particular value $q=1/s$. In this case we impose that each 1--peak of a Dyck path $P$ must be followed by at least $s$ consecutive 0--valleys if after the $1$--peak there is enough space to contain $s$ consecutive $0$--valleys. If a $1$--peak occurs near the end of $P$ and there is no space to contain $s$ consecutive $0$--valleys, then no other $1$--peaks can occur up to the end of $P$. Summarizing, we give the following definition.

\begin{definition}\label{1 su s}
Let $\mathcal D_n^{1/s}$ denote the set of Dyck paths $P$ of semilength $n$ having height at most $2$, where either $P$ has no $1$--peaks ($P=(UD)^n$) or each $1$--peak in $P$ is followed by at least $s$ consecutive $0$--valleys, except the last $1$--peak which can be followed by less then $s$ consecutive $0$--valleys.
\end{definition}

The construction is straightforward: a path $P\in\mathcal D_n^{1/s}$ starts with one of the $UD$ or $UUD(DU)^{s-1}D$ factors (or $P$ is a suitable prefix of this last one factor, if $n\leq s$), then $P$ is obtained by concatenating one of these factor to a Dyck path $Q\in\mathcal D_{n-1}^{1/s}$ or $Q\in\mathcal D_{n-s-1}^{1/s}$. 
In the case $P$ begins with the longer factor, since $Q$ starts, of course, with an up step $U$, then really the first $1$--peak of $P$ is followed by $s$ consecutive $0$-valleys (so that $P\in\mathcal D_n^{1/s}$): the first $s-1$ ones are the $s-1$ consecutive $0$--valleys of the factor, and the last one is given by the last step $D$ of the factor and the first $U$ step of $Q$.

Not considering, for the moment, paths with a semilength less or equal to $s$, we can write
\begin{equation}\label{construction}
\mathcal D_n^{1/s} = 
UD\cdot \mathcal D_{n-1}^{1/s}\ \cup\ UUD(DU)^{s-1}D\cdot \mathcal D_{n-s-1}^{1/s}\ (\text{for $n>s$}).
\end{equation}
Thus, the set $\mathcal D_n^{1/s}$ is enumerated by
$$
w_n=w_{n-1}+w_{n-s-1}
$$
omitting at this stage the initial conditions.

We note that this recurrence relation matches the one enumerating the $q$-decreasing strings in the case $q=1/s$ stated in \cite{Kir}.

As far as the initial conditions are concerned, we observe that prepending the factor $UD$ generates Dyck paths having semilength starting from $1$, so that the empty Dyck path $\varepsilon$ must be considered as a legal path of $\mathcal D_0^{1/s}$ (actually, the only one!).

Prepending the factor $UUD(DU)^{s-1}D$ generates Dyck paths having semilength starting from $s+1$. For semilengths less or equal to $s$, 
we note that the construction described in equation (\ref{construction}) does not generate the paths $UUDD$, $UUDDUD$, $\ldots$, $UUD(DU)^{t}D$ with $t=0,1,\ldots, s-2$. These paths are suitable prefixes of $UUD(DU)^{s-1}D$ which however satisfy Definition \ref{1 su s}, so that they must be considered among the initial conditions.

Therefore, the generation of the set $\mathcal D_n^{1/s}$ can be completely described as follows:

$$
\mathcal D_n^{1/s}=
\begin{cases}
\varepsilon,  & \text{if $n=0$};\\
UD, & \text{if $n=1$};\\
UD\cdot \mathcal D_{n-1}^{1/s}\ \cup\ U \cdot p_{n-1}\left(UD(DU)^{s-1}\right) \cdot D,& \text{if $2\leq n\leq s+1$};\\
UD\cdot \mathcal D_{n-1}^{1/s}\ \cup\ UUD(DU)^{s-1}D\cdot \mathcal D_{n-s-1}^{1/s},&\text{if $n>s+1$}.
\end{cases}\quad 
$$
In the above formula $p_{n-1}\left(UD(DU)^{s-1}\right)$ is the prefix of semilength $n-1$ of $UD(DU)^{s-1}$.

It is not difficult to see that $\mathcal D_n^{1/s}$ is enumerated by

\begin{equation*}
w_n=
\begin{cases}
1,  & \text{if $n=0$};\\
n, & \text{if $1\leq n \leq s+1$};\\
w_{n-1}+w_{n-s-1}, & \text{if $n > s+1$}.\\
\end{cases} 
\end{equation*}
This sequence matches the one enumerating the $q$-decreasing strings for $q=1/s$ that can be deduced from \cite{Kir}.

\subsection{The general case $q=r/s$}
Following the outline of the constructions in the cases where $q$ is an
integer, and where $q=1/s$, in the general case $q=r/s$ (we suppose $r$ and $s$ to be coprime) we require that a path $P$ avoid $r+1$ consecutive $1$--peaks, and if $r$ consecutive $1$--peaks occur in $P$, then they must be followed by at least $s$ consecutive $0$--valleys. Clearly, we have to deal with the case where $p$ consecutive $1$--peaks, with $p=1,2,\ldots,r-1$, occur in $P$. When this happens, we impose that the $p$ consecutive $1$--peaks must be followed by a number $v$ of consecutive $0$--valleys such that

\begin{equation}\label{numvalli}
\frac{p}{v}\leq \frac{r}{s}\quad\quad \text{for $p=1,2,\ldots,r-1$} .
\end{equation}
Moreover, we have to deal with the case where the rightmost block of $r$ consecutive $1$--peaks occurs near the end of $P$ (more precisely, when after this block there is no space to contain $s$ consecutive $0$ --valleys). In this case, no other $1$-peaks can occur up to the end of $P$. Finally, we allow the paths end with a consecutive block of $1$--peaks (clearly, less than $r+1$). Summarizing, we give the following definition.

\begin{definition}\label{r su s}
Let $\mathcal D_n^{r/s}$ denote the set of Dyck paths $P$ of semilength $n$ having height at most $2$, where
\begin{itemize}
    \item each block $B$ of $r$ consecutive $1$--peaks in $P$ is followed by at least $s$ consecutive $0$--valleys, except the rightmost block $B$ which can be followed by less than $s$ consecutive $0$--valleys, and
    \item each block $C$ of $p$ consecutive $1$--peaks in $P$, with $p=1,2,\ldots,r-1$, is followed by at least $v$ consecutive $0$--valleys such that $p/v\leq r/s$, and
    \item the path P can end with $(UD)^tD$, with $t=1,2,\ldots,r$ (in other words, P ends with $t$ consecutive $1$--peaks ($t\leq r$) followed by a down step).
\end{itemize}
\end{definition}
Clearly, the number $v$ of consecutive $0$--valleys, since $v$ is an integer, satisfies our request (\ref{numvalli})
 when
$
v\geq\left\lceil{p\cdot\frac{s}{r}}\right\rceil\ .
$
Moreover, we note that in the case where $p$ consecutive $1$-peaks, with $p=1,2,\ldots,r-1$, occur near the end of $P$ and there is no enough space to contain $v$ consecutive $0$-valleys, then the path $P$ does not belong to $\mathcal D_n^{r/s}$, according to the second bullet in Definition (\ref{r su s}). For example, if $q=4/5$, the path $P=UUDUDDUDUD$ is not allowed, since after two consecutive $1$--peaks ($p=2$) at least three consecutive $0$-valleys ($v=3$, according to request (\ref{numvalli}))
 must occur. On the other hand, the path $P=UUDUDUDUDDUDUDUD$ is allowed.

Also in this (general) case, the construction is straightforward.  
A path $P\in\mathcal D_n^{r/s}$ starts with one of the factors $UD$ or $U(UD)^{p}(DU)^{\lceil{p s/r}\rceil-1}D$, with $p=1,2,\ldots,r$ (or $P$ is a suitable prefix of $U(UD)^{r}(DU)^{s-1}D$). 
With an argument similar to the one used in the case where $q=1/s$, it is not difficult to get the following construction:

$$
\mathcal D_n^{r/s}=
\begin{cases}
\varepsilon,  & \text{if $n=0$};\\
&\\
UD, & \text{if $n=1$};\\
&\\
UD\cdot \mathcal D_{n-1}^{r/s}\ \cup\ 
U\cdot p_{n-1}\left((UD)^{r}(DU)^{s-1}\right)\cdot D& \\
\cup \ U(UD)^{p}(DU)^{\lceil{p s/r}\rceil-1}D\cdot \mathcal D_{n-p-\lceil{p s/r}\rceil}^{r/s}, & \text{if $2\leq n \leq r+s$}\\
&\text{and $n - p -\lceil{ p s/r}\rceil \geq 1$}\\ 
&\text{with $1 \leq p \leq r-1$};\\
&\\
UD\cdot \mathcal D_{n-1}^{r/s}\ \cup\ 
U(UD)^{p}(DU)^{\lceil{p s/r}\rceil-1}D\cdot \mathcal D_{n-p-\lceil{p s/r}\rceil}^{r/s},&\text{if $n > r+s$}\\ 
& \text{and $1 \leq p \leq r$;}
\end{cases}\quad 
$$
Then we have (in the following $\chi(f)=1$ if $f$ is true, and $\chi(f)=0$ otherwise)

$$
w_n=
\begin{cases}
1,  & \text{if $n=0$};\\
1,  & \text{if $n=1$};\\
w_{n-1} + 1 + \displaystyle\sum_{p=1}^{r-1} \chi \left( n- p -
\lceil{p s/r}\rceil \geq 1 \right) w_{n-p-\lceil{p s/r}\rceil},& \text{if $2\leq n \leq r+s$};\\
w_{n-1} + \displaystyle\sum_{p=1}^{r} w_{n-p-\lceil{p s/r}\rceil}, & \text{if $n > r+s$}.\\
\end{cases} 
$$

In order to respect the second bullet in Definition \ref{r su s} we added the factor $\chi \left( n- p -
\lceil{p s/r}\rceil \geq 1 \right)$ in the case $2\leq n \leq r+s$ of the definition of $w_n$, and the statement $n- p -
\lceil{p s/r}\rceil \geq 1$ since at least $\lceil{p s/r}\rceil$ consecutive $0$--valleys must occur after $p$ consecutive $1$--peaks, with $1 \leq p \leq r-1$ (in order to have $\mathcal D_{n-p-\lceil{p s/r}\rceil}^{r/s} \neq \emptyset$).

Also, in this case, the recurrence relations match the ones enumerating the $q$-decreasing strings in the case $q=r/s$ founded in \cite{Kir}.

\section{Acknowledgements}

This work is partially supported by the INdAM -- GNCS Project 2023 ``Aspetti combinatori ed enumerativi di strutture discrete: stringhe, ipergrafi e permutazioni'', code CUP\_ E53C22001930001.

\nocite{*}
\bibliographystyle{eptcs}
\bibliography{generic}

\end{document}